\documentclass{smarthepnote}
\usepackage[colorinlistoftodos]{todonotes}
\usepackage{placeins}
\usepackage{titlesec}

\usepackage{dirtytalk}
\usepackage{siunitx}
\usepackage{subcaption}

\title{\vspace{-0.5cm}Summary of the trigger systems of the Large Hadron Collider experiments ALICE, ATLAS, CMS and LHCb\vspace{-0.25cm}}
\author{The SMARTHEP Network\vspace{-1cm}}
\date{\today}

\DocAuthors{V.~Gligorov \\ C.~Doglioni \\ the SMARTHEP network}
\DocEditors{\vspace{-0.25cm} L. Bozianu \\ S. Cella \\ C. E. Cocha Toapaxi \\ K. Endrup Iversen (*) \\ P. Inkaew \\ D. Magdalinski \\ D. J. Wilson-Edwards (*)\vspace{-0.25cm}}
\DocCoordinators{J. Albrecht \\ L. Calefice \\ J. A. Gooding \\ A. Sopasakis }
\DeliverableNo{6.1}
\draftversion{3.0 (Preprint for arXiv)}

\def\thickhline{\noalign{\hrule height1.2pt}}

\begin{document}
\maketitle

\begin{abstract}
%This white paper from the SMARTHEP Network reviews the current status of trigger systems employed at the Large Hadron Collider (LHC). The principles of triggering in the current state-of-the-art are discussed and illustrated with recent and ongoing developments from the ALICE, ATLAS, CMS and LHCb collaborations. This review places particular emphasis on real-time data processing approaches and techniques.

\vspace{-0.25cm}
In modern High Energy Physics (HEP) experiments, triggers perform the important task of selecting, in real time, the data to be recorded and saved for physics analyses. As a result, trigger strategies play a key role in extracting relevant information from the vast streams of data produced at facilities like the Large Hadron Collider (LHC). As the energy and luminosity of the collisions increase, these strategies must be upgraded and maintained to suit the experimental needs. This whitepaper compiled by the SMARTHEP Early Stage Researchers presents a high-level overview and reviews recent developments of triggering practices employed at the LHC. The general trigger principles applied at modern HEP experiments are highlighted, with specific reference to the current trigger state-of-the-art within the ALICE, ATLAS, CMS and LHCb collaborations. Furthermore, a brief synopsis of the new trigger paradigm required by the upcoming high-luminosity upgrade of the LHC is provided. 

\textit{This white paper is not meant to provide an exhaustive review or substitute documentation and papers from the collaborations themselves, but rather offer general considerations and examples from the literature that are relevant to the SMARTHEP network.} 

%the application of machine learning techniques for real-time analysis (RTA) of collision events at the Large Hadron Collider (LHC). It discusses the crucial role of RTA in addressing the data-processing challenges faced by LHC experiments and how modern ML techniques are helping to overcome them. First, we provide some insights into selecting the most suitable ML algorithms and adapting them to the real- time environment using software techniques and hardware accelerators. Then, we review a small selection of current and future uses of machine learning for RTA that are of particular interest to the SMARTHEP network. We continue by emphasizing the importance of collaborations between the high-energy physics community and industry in tackling these challenges and we showcase the mutual benefits that can be achieved. 
%The white paper is not meant to provide an exhaustive review, but rather some general considerations and a limited number of relevant examples from the literature that are relevant to the SMARTHEP innovative training network

%This document is a draft of the trigger whitepaper written by the ESRs that is going to be circulated within the SMARTHEP network. Parts of it may be made public at a later stage. It is intended to serve as an exercise for the SMARTHEP ESRs to learn more about the trigger systems of different experiments (since many ESRs chose to write about the trigger systems of experiments that they are not members of). 
\vspace{-0.25cm}
\end{abstract}

% Make the review table at the bottom of the title page
\vspace{-0.25cm} %\vfill
\makereviewtable
\clearpage

% Short documentes dont always need a Table of Content / Figures / Tables, so comment out what is not needed
\begingroup
\color{black}
%\tableofcontents
%\listoffigures
%\listoftables
\endgroup
\pagebreak

\section{Introduction}

%Intro

The Large Hadron Collider (LHC) was designed to operate at the highest possible energy and luminosity~\cite{CERN:lhc-design-report} to enable precise measurements of the fundamental components of matter and their interaction, and to seek new physics phenomena. 
In the latest LHC data-taking period, Run 3, protons are accelerated to an energy of \SI{6.8}{\tera\electronvolt}, grouped into bunches in opposing beams which cross one another every \SI{25}{\nano\second}~\cite{CERN:lhc-run3-operation}. 
Capturing the details of proton-proton ($pp$) collisions at a rate of \SI{40}{\mega\hertz} introduces an immense data challenge, in which recording collisions in full detail at a typical LHC experiment would require transferring and writing data at up to ${\sim}\SI{40}{\tera\byte\per\second}$. 
To effectively handle this data flow, High Energy Physics (HEP) experiments employ trigger and data acquisition systems, designed to perform detector data readout, event-building, selection, etc in near-real time, reducing the throughput to a manageable level. Typically, $\mathcal{O}\left(\SI{1}{\giga\byte\per\second}\right)$ of data useful to the physics goals of an experiment is written to permanent storage.

Each LHC experiment discussed in this paper — ALICE, ATLAS, CMS and LHCb — has undertaken considerable research and development in terms of triggers and data acquisition (TDAQ), and their most recent reviews can be found in Refs.~\cite{alice-performance-paper-run1, ATLASTriggerRun3, CMS:run3-detector, LHCb:upgrade_trigger_TDR}.

Developments in computational resources (e.g., the adoption of hybrid architectures) and data processing approaches (e.g., real-time and parallelised software frameworks) have enabled more advanced trigger systems to be developed in recent years~\cite{GPUs-for-hep}. These trigger systems are capable of performing real-time calibration, high-level monitoring, etc. ALICE and LHCb performed significant upgrades to their trigger systems ahead of Run~3 of the LHC (2022-2025)~\cite{alice-upgrade, LHCb:2023hlw}, with ATLAS and CMS planning upgrades of similar scales ahead of the High-Luminosity LHC (HL-LHC) operational period (2029-2040s)~\cite{ATLAS:upgrade-scoping, CMS:upgrade-hllhc}.
In this paper, the current trigger state-of-the art is reviewed, predominantly within the context of Run~3.

% Overview of trigger strategies
% Model detector, what does a non-specific trigger look like?
\subsection{Principles of triggering in High Energy Physics}

\begin{figure}[!ht]
    \centering
    \includegraphics[width=\linewidth]{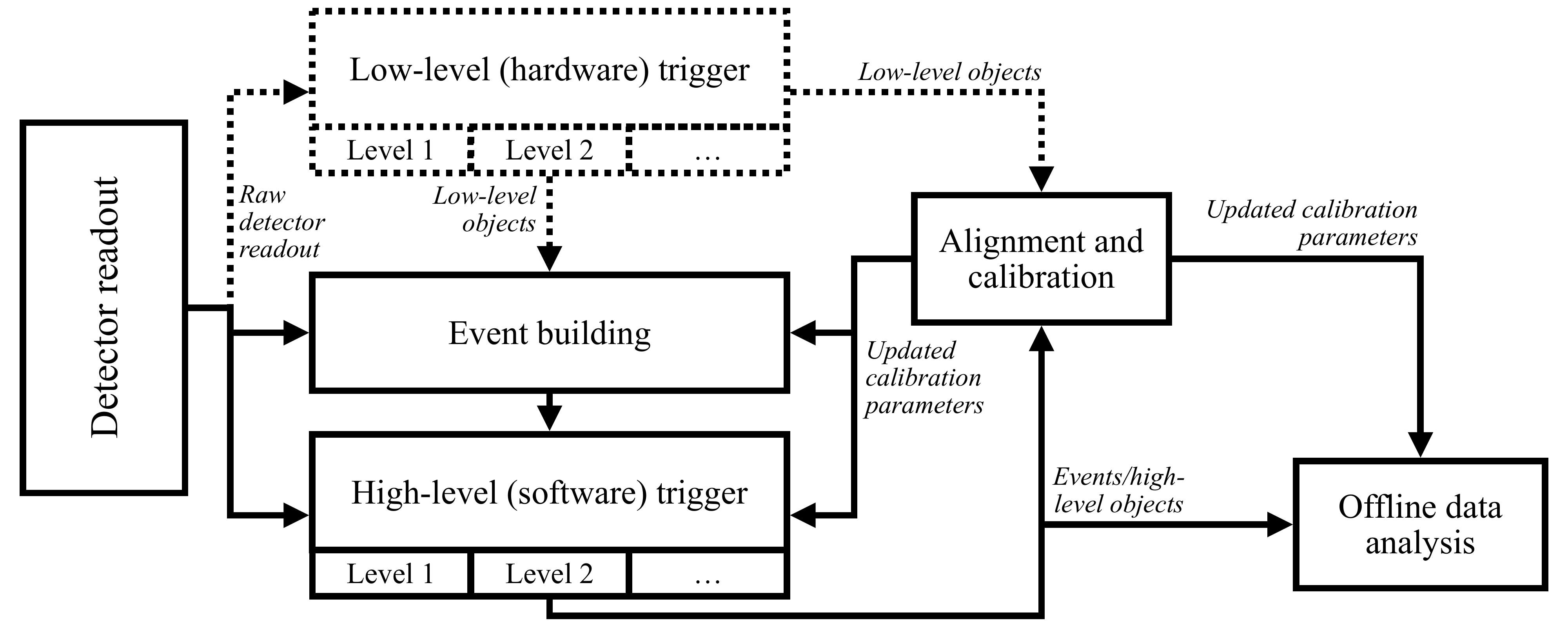}
    \caption[A sketch of a trigger system for a  HEP experiment in the current paradigm. Raw data is read out from each subdetector. A low-level (hardware) trigger is typically employed to synchronise readout and apply initial selection/compression algorithms. A high-level (software) trigger makes further selections on reconstructed objects of increasing accuracy, which are then saved for offline analysis. Alignment and calibration are often performed alongside the high-level trigger to inform the reconstruction algorithms required for detailed selections.]{A generalised schematic of a trigger system for an HEP experiment in the current paradigm. Raw data is read out from each subdetector. A lower-level (hardware) trigger is typically employed to synchronise readout and apply initial selection/compression algorithms. Low-level trigger information and detector readout/reconstructed objects are combined in an event building stage. A high-level (software) trigger makes further selections on reconstructed objects of increasing accuracy, which are then saved for offline analysis. Alignment and calibration are often performed alongside the high-level trigger to inform the reconstruction algorithms required for detailed selections.}
    \label{trigger-schema}
\end{figure}

The primary function of a trigger system, as laid out in Figure~\ref{trigger-schema}, is to reduce the data rate to be processed, according to the physics priorities of the  experiment~\cite{Jeitler_2017, Smith2020, Beck_2007}. 
The performance of a trigger system can therefore be optimised according to three quantities: high signal efficiency, high background rejection (or equivalently low background efficiency) and affordable throughput/output bandwidth. 
Furthermore, this performance must be quantifiable (e.g., that trigger efficiencies are calculable), robust and deterministic. 
To satisfy these requirements, triggers are typically designed in a tiered structure: a lower-level (often hardware-based) tier performing initial data reduction and coarse selection; followed by a software-based (high-level) tier performs reconstruction of physics objects upon which further selections of events is performed. 
This is not the case for all LHC experiments: LHCb does not employ a lower-level hardware trigger in Run~3~\cite{LHCb:upgrade_trigger_TDR} and the ALICE lower-level trigger does not apply selection~\cite{alice-trigger-run3}.

A lower-level trigger must operate at a rate sufficiently close to the collision rate, and thus must have a minimal dead time between operations. 
The lower-level trigger operates in a staged configuration, increasing latency with each stage to enable more detailed (and thus slower) operations to be performed on the data. 
They are often implemented in custom high-speed electronics to perform operations meeting the latency requirements. 
The data processed by the low-level trigger, typically reduced in rate by a factor $\mathcal{O}\left(100\right)$, is then passed to the high-level trigger for more detailed processing.

Higher-level triggers perform more complex and computationally intensive tasks such as event reconstruction and calibration. 
To ensure that such tasks can be performed at the low-level trigger output rate, the high-level trigger can be separated into two tiers, with the first performing coarse reconstruction and initial selections requiring simpler reconstructed objects (e.g., tracks), and the second performing a detailed reconstruction for more complex selections. 
High-level triggers typically are implemented in computing farm, making use of parallel computing approaches or hybrid computing architectures to efficiently perform tasks within the available computing resources.

The implementations of such systems in each of the major LHC experiments are introduced in the following sections. 

%It is difficult to satisfy both the high efficiencies and high background rejection physics requirements, as well as the very small trigger latency to reduce dead time. Multi-trigger systems are usually introduced. The Level-1 (L1) Trigger is a hardware trigger, comprised of custom high-speed electronics, which makes decisions based off of coarsely reconstructed subset of information from the detector. The L1 trigger renders a selection decision every bunch crossing, with high signal efficiency and comparatively lower background rejection. The data harvested from the L1 trigger is typically more than experiment infrastructure can handle for permanent storage and offline data analysis. Therefore, more sophisticated software Higher Level Triggers (HLT) are usually implemented to further reduce the rate, using full-precision and finer granularity detector information. The maximum allowable acceptance rate L1 trigger is constrained by the detector read out, the speed of at which the HLT performs the more refined selection, and the rate at which the Data AcQuisition (DAQ) system can retrieve the data for permanent storage.  

%The DAQ bandwidth, which is determined by the available storage capabilities and computing power, is related to the maximum allowable trigger rate $R_{\mathrm{trigger}} ^ {max}$ as follows:

\subsection{Trigger systems of the ATLAS and CMS experiments}

The two general-purpose LHC experiments, ATLAS and CMS, have similar, broad physics programmes~\cite{ATLASMachine,collaboration2008cms}. 
These programmes include both measurements of the Standard Model and searches for new physics: indirectly, by probing the nature of known particles at the precision frontier, and directly (e.g. through searches for new particles and dark matter candidates~\cite{snowmass-darkmatter}). 
Both experiments consist of concentric layered subdetectors; namely inner detectors for charged particle tracking, calorimeters for measurement of hadron/electromagnetically interacting particle energies and positions, and muon spectrometers for the identification and precise measurement of muons. 
Due to the large acceptance and high granularity of the subdetectors used, the experiments contain a large number of detector channels, resulting in event sizes of the order of  ${\sim}\SI{1}{\mega\byte}$. 
ATLAS and CMS implement similar two-tier trigger systems with a Level-1 (L1) hardware-based trigger and a software-based high-level trigger (HLT). In Run~2 and Run~3, the triggers of both experiments reduced the initial \SI{40}{\mega\hertz} bunch crossing rate to an L1 acceptance rate of \SI{100}{\kilo\hertz}, reduced further to a final HLT output rate of ${\sim}\SI{1}{\kilo\hertz}$. For average event sizes ranging from ${\sim}\SI{500}{\kilo\byte}$~\cite{ATLASRun3EventBuilder} to ${\sim}\SI{2}{\mega\byte}$~\cite{cmsRun3EventBuilder}, this corresponds to a HLT output bandwidth of $\mathcal{O}\left(\SI{1}{\giga\byte\per\second}\right)$.

\subsection{Trigger system of the LHCb experiment}

\begin{figure}[h!]
    \centering
    \includegraphics[width=0.55\linewidth]{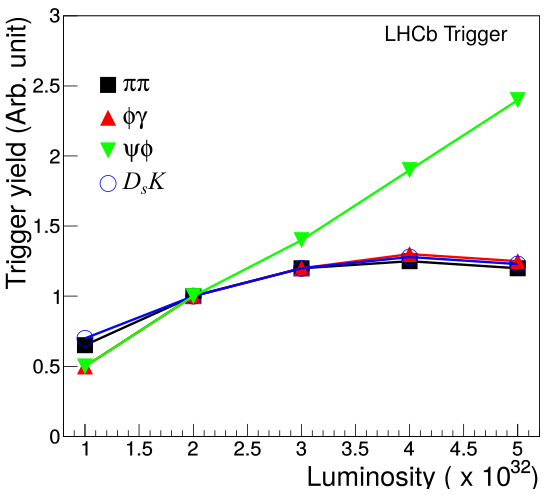}
    \caption{Trigger yield as function of luminosity (in ${\rm cm}^{-2}{\rm s}^{-1}$) shown relative to the yield at $2\times 10^{32}~{\rm cm}^{-2}{\rm s}^{-1}$ with the Run 2 trigger configuration from Ref.~\cite{LHCb:upgrade-piucci}. Yields are shown for four $B$ meson decay modes representative of the LHCb $b$ physics programme. Only the decay mode $\psi\phi$ observes an increase in luminosity from accelerator upgrades. In the remaining modes this increase is suppressed by the efficiency of the L0 trigger.}
    \label{fig:LHCbL0TriggerYield}
\end{figure}

The LHCb experiment is operating in the forward region of LHC collisions and was originally designed to precisely study heavy-flavour decays, including those involving $CP$~and~flavour violation~\cite{LHCb:detector-paper}.
Currently, the LHCb detector operates as a general-purpose forward detector with a broad physics program including e.g. dark matter searches and electroweak physics. 
The LHCb trigger system was redesigned for Run~3, removing the low-level Level 0 (L0) hardware-based trigger previously employed in Runs 1 and 2~\cite{LHCb:upgrade_trigger_TDR}.
At the Run~3 LHCb luminosity of about \SI{2e33}{\per\square\cm\per\second}, the maximum output rate of \SI{1}{\mega\hertz} of the L0 trigger and its simple selection criteria based on transverse momenta would result in a saturation of the yields of modes of interest. 
This is demonstrated in Figure~\ref{fig:LHCbL0TriggerYield}, wherein the yields of processes involving light hadrons ($\pi\pi$, $D_s K$) and photons ($\phi\gamma$) plateau with increasing luminosity.

The upgraded LHCb trigger consists solely of a High-Level Trigger (HLT), split between two stages: HLT1 and HLT2~\cite{Aaij:2019uij}. Following the removal of the L0 trigger, LHCb event readout has increased from \SI{1}{\mega\hertz} to \SI{30}{\mega\hertz}\footnote{The LHCb Event Builder was designed to handle the average bunch crossing of 30 MHz, as well as the maximum crossing rate of 40 MHz.}. During Run~2, HLT1 and HLT2 were decoupled to allow HLT1 to run synchronous to data-taking and HLT2 to run asynchronously, enabling detector calibrations between the two steps to improve reconstruction performance to a quality previously only achieved offline~\cite{LHCb:upgrade_trigger_TDR}.

HLT1 was upgraded to perform a partial reconstruction of the full detector readout. To achieve this, the reconstruction algorithms were upgraded to run on GPUs hosted on the same event-building servers that host the FPGA cards required to receive data from the detector at \SI{30}{\mega\hertz}~\cite{LHCb_Allen_GPU}. Reconstructed, selected events are propagated to a buffer at an event rate of ${\sim}\SI{1}{\mega\hertz}$. HLT2, implemented as a CPU farm known as the Event Filter Farm (EFF), takes as input the most recent detector alignment and calibration, reconstructing events in full offline quality for detailed selection. This selection reduces the output bandwidth to \SI{10}{\giga\byte\per\second}, corresponding to an event rate of $\sim\SI{100}{\kilo\hertz}$~\cite{lhcb_hlt2_storage_run3}.

\subsection{Trigger system of the ALICE experiment}
The ALICE experiment is dedicated to the study of heavy-ion collisions at the LHC, with a focus on studies of quantum chromodynamics in energy-dense environments (e.g.,  quark-gluon plasma)~\cite{alice-performance-paper-run1}. To study such environments, ALICE studies p-p, p-Pb and Pb-Pb collisions at rates of \SI{1}{\mega\hertz}, \SI{500}{\kilo\hertz} and \SI{50}{\kilo\hertz}, respectively~\cite{alice-trigger-run3}. Heavy ion collisions result in a very high multiplicity of particles, with ${\sim}\SI{700}{\mega\byte}$ of raw data per collision event collected by the ALICE experiment~\cite{alice-rta-trigger}. The ALICE detector is barrel-shaped, containing concentric particle tracking and identification systems, and a forward muon spectrometer. At the core of the barrel is a Time Projection Chamber (TPC) vital to tracking performance, contributing the majority of the event size (\SI{95.3}{\percent} and \SI{91.1}{\percent} of total data volume in Pb-Pb and p-p collisions, respectively).

The ALICE trigger was upgraded ahead of Run~3 to facilitate continuous readout of subdetectors at the collision rates listed above. The Central Trigger System (CTS), is responsible for the synchronisation of raw detector readout. The CTS transmits aggregated data and trigger signals to the HLT. The HLT then performs event reconstruction, data volume reduction and subdetector calibration, processing data at a maximum rate of \SI{48}{\giga\byte\per\second} to achieve an output throughput of up to \SI{12}{\giga\byte\per\second}~\cite{alice-rta-trigger}.

\section{Data processing and selections in the lower-level triggers}
% Introduce and group ATLAS+CMS+ALICE/LHCb
The first stage of any trigger system requires the readout of reduced granular detector information (e.g., energy deposits in a calorimeter/tracker) and processing of this information. 
This processing has two main steps. First, information from the subdetectors must be synchronised to ensure that the corresponding data are correctly associated to an event.  
Subsequently, the data volume needs to be reduced through a combination of basic compression techniques (e.g., zero suppression) and selections. 
This is the common task of a low-level trigger - once the event is selected, the detector (or subdetector) is read out. 
This section covers only ATLAS, CMS and ALICE since LHCb no longer employs a low-level trigger: detector information is continuously read out (with compression applied upon readout) and passed to the event builder network of HLT1~\cite{LHCb:2023hlw}.
This approach is better suited to LHCb, with smaller event sizes and a non-hermetic detector geometry (readout cables sit outside detector acceptance).

%The smaller event size and non-hermetic layout(readout cables sit outside detector acceptance) means this change is more achievable for LHCb than ATLAS and CMS.
%\todo[inline]{Waiting for feedback on the sentence above} 
%Large ATLAS on tiered trig sys, similar structure in CMS, slight difference in ALICE, then tiny LHCb caveat/sentence to be discussed later

%ATLAS+CMS+ALICE
The initial hardware-based L1 triggers of ATLAS~\cite{ATLASRun3Detector} and CMS~\cite{CMS:run3-detector} both reduce the rate of data down to a maximum detector read out limit of \SI{100}{\kilo\hertz}, within a latency of \SI{2.5}{\micro\second} at ATLAS and \SI{4}{\micro\second} at CMS. 
The systems consist of custom Algorithm Specific Integrated Chips (ASIC)- and Field Programmable Gate Arrays (FPGA)-based electronics~\cite{asics-fpgas}. 
For Run 3, the ATLAS and CMS L1 triggers use reduced granular information from the calorimeter and muon systems (and the combination of the two) to perform coarse selections~\cite{ATLASTriggerRun3, cms2020performance}. 
Upon an event passing this selection, the L1 directs the readout hardware of each subdetector to process the surviving data, which is stored in associated local buffers. 
Information from each subdetector is  then readout, processed and combined within a central readout system where events are assembled. 
It is buffered in the HLT farm until requested for processing by the HLT. 
The data acquisition (DAQ) systems of both experiments have evolved to use consumer network and computing hardware downstream of custom on-board electronics. 
This setup simplifies the readout in complexity, maintenance cost and upgrade capabilities. 
%At ATLAS the Front-End LInk eXchange (FELIX)~\cite{ATLAS:FELIX} readout boards, responsible for the interface between commercial and custom hardware, have been partially implemented in Run 3 and will be fully implemented for HL-LHC.
%\todo[inline]{Consider removing the above sentence because we don't have equivalents for ALICE and CMS}

In Run~3, several of the ALICE subdetectors have been upgraded to read out data continuously. The CTS synchronises data, subdividing readout into HeartBeat (HB) frames of approximately 1 LHC orbit period ($\sim\SI{88.92}{\micro\second}$). An HB frame is only kept if all relevant readout units (up to 441 in the entire detector) can be read out. Each HB decision is transmitted asynchronously to the First Level Processor, instructing it on what data to keep in a given HB frame. Whilst many subdetectors (including the TPC) were upgraded to read out data continuously in Run~3, continuous readout is not possible for some subdetectors, e.g.,  the Transition Radiation Detector. Such subdetectors are operated on a triggered basis and are hence excluded from the HB decision calculation, instead using the existing RD12 TTC protocol developed for Run~2 operation. Triggered subdetectors thus operate independently, with triggered data combined with continuous readout data at a later stage~\cite{alice-trigger-run3}.

\section{Reconstruction of physics objects in the higher-level triggers}

The reconstruction of events from detector information is a key function of any trigger system; trigger selections typically involve imposing conditions on reconstructed objects including particle tracks and calorimeter clusters. 

In the following, we will focus on examples of reconstruction tasks that are performed \textit{online} within the trigger system of different experiments. 
We denote with \textit{offline} reconstruction tasks those that are performed on objects in events selected by the trigger system at a later stage.  

The complexity and performance requirements of online reconstruction methods can range from fast, lower-level algorithms to full, offline-quality reconstruction of entire events and is determined by the requirements of the given trigger stage. 
It is important for online algorithms to be aligned with (ideally identical to) offline algorithms in terms of inputs, implementation and performance, so that there is a good match between physics objects that inform the event selection and physics objects used for analysis. 
This is also crucial for real-time physics analysis, as discussed in Section \ref{sec:RTA_physics}. 

We refer to Refs.~\cite{alice-upgrade, ATLASTriggerRun3, CMS:run3-detector, LHCb:upgrade_trigger_TDR} for a full description of the algorithms used by the different experiments.

\subsection{The LHCb online tracking algorithms}

In this section we describe the LHCb track reconstruction procedure as an example of the processes involved in reconstructing event objects such as tracks. Similar processes take place in the reconstruction chains of the other LHC experiments with variations due to the specific detector architecture.
% The LHCb reconstruction procedure provides a clear example of the processes involved in reconstructing events and event objects. 

In HLT1, information from the three tracking detectors—Vertex Locator (VELO)~\cite{LHCb:velo-tdr}, Upstream Tracker (UT) and Scintillating Fibre (SciFi) Tracker~\cite{LHCb:tracker-tdr} are used to reconstruct partial event objects (e.g., particle tracks) in the dedicated CUDA-based software framework Allen~\cite{LHCb_Allen_GPU}. 
%\todo[inline]{Need to define partial event reconstruction}
For each sub-detector, many of the following processes must be performed: decoding of input into the global coordinate system; clustering of hits within a detection plane; combination of hits/clusters from different detection layers to form particle trajectories; fitting of track model to track candidates; vertex-finding between tracks.

For the VELO, reconstruction begins with the clustering of hits on each silicon plane, with a bitmask-based clustering algorithm operating in parallel across the local regions of each cluster. Straight-line tracks are then reconstructed, starting with seeds of three hits from consecutive silicon planes, extended to the remaining layers and fit with a simple Kalman filter~\cite{kalman-track}. 
Finally, primary vertex (PV) candidates are identified and matched to the tracks. Rather than mapping tracks one-to-one with PV candidates, each track receives a per-candidate weight (corresponding to the likelihood it is associated to each PV candidate) to enable parallel computation. 

UT hits are instead assigned to extrapolated VELO tracks using a minimum momentum cut-off, with the track momentum calculated from the curvature between the straight-line VELO tracks and UT hits. 

Tracks passing the VELO and UT are extrapolated to create a search window in the SciFi, where seeds of three hits in different layers are formed. A $\chi^2$ fit is performed on the seeds, and the best seeds are extrapolated to the remainder of the SciFi layers. Since the discrimination power of the three initial hits is limited, several seeds are extrapolated for each UT track, performing additional fits to select the best track per UT track. Hits in the muon systems are matched to extrapolated SciFi tracks according to the track parameters obtained in the previous steps~\cite{LHCb:2023hlw, LHCb_Allen_GPU}.

These algorithms are applied per-track in parallel on the GPUs in the event-building servers. The high-throughput provided by this mass-parallelisation thus enables HLT1 to perform reconstruction of tracks at the full readout rate of \SI{30}{\mega\hertz}~\cite{LHCb_Allen_GPU}.

% \todo[inline]{Not sure how the paragraph below fits in here, consider removing?}
% Another class of algorithms that is widely used at LHC experiments is Kalman filters (KFs). KFs are recursive algorithms used for state estimation and data fusion (i.e., combination of information from multiple sensors), with a well-established position as a key reconstruction tool~\cite{kalman-track}, and are particularly well-suited for problems involving dynamic (i.e.,  time-dependent) systems and noisy measurements. KFs combine information from previous state estimates and new measurements to provide an optimal estimate of the current state, taking into account both the dynamics of the system and the uncertainties associated with each measurement~\cite{FRUHWIRTH1987444}. KFs are employed across the major LHC experiments for the tracking of charged particles~\cite{Belikov:2003yr,ATLAS:tracking,CMS:tracking,LHCb_Allen_GPU}

\subsection{The CMS Particle Flow algorithm} \label{sec:Algorithms_PFlow}

An ever-increasing collection of algorithms for reconstruction of events and constituent objects are developed and adopted by collider physics experiments. 
One example is Particle Flow (PF)~\cite{sirunyan2017pflowcms} that combines measurements from the various sub-detectors in the ATLAS and CMS sub-detectors to produce particle candidates from the entire event. The general PF process can be briefly summarised as follows: the full detector output is used to describe the global collision event, identifying several basic elements individually and iteratively clustering them together into more complex composite physics objects~\cite{CMS:2020uim,CMS:2018rym,CMS:2014pgm}. 
These basic elements are used to build electrons, muons, tau leptons, photons, jets, missing transverse momentum and other physics objects~\cite{CMS:2018jrd,CMS:2016lmd,CMS:2019ctu}. 
The physics performance is ameliorated by the combination of measurements from the different sub-detectors which achieve optimal accuracy in different regions of phase space. 
For example, the best momentum resolution is attained at low transverse momenta in the inner detector tracker and at high transverse momenta in the calorimeter. The composition of the tracks and calorimeter clusters (particle flow objects (PFOs)) are used in jet reconstruction to produce particle flow jets.

The PF procedure is widely-used in CMS reconstruction in both the trigger and offline analyses. The CMS detector design is well-suited to the use of PF reconstruction: a highly-segmented tracker, fine-grained electromagnetic calorimeter, hermetic hadron calorimeter, strong magnetic field and excellent muon spectrometer provide full, high quality coverage~\cite{sirunyan2017pflowcms}. The ATLAS experiment has expanded its usage of PF in the trigger for Run 3 to include hadronic jet reconstruction. The improvement in resolution is less pronounced in ATLAS due to the good energy resolution of the calorimeters and jets produced using only calorimeter clusters, however there is significant improvement in the rejection of jets from collisions occurring in addition to the collision of interest within the same detector sensitivity window (pile-up)~\cite{ATLASTriggerRun3,ATLASJetPFlow}. 

% a widely-used reconstruction procedure in CMS offline analyses, now considered the ``baseline" for object reconstruction in the CMS HLT. The purpose of the PF algorithm is to reconstruct and identify all particles involved in a collision by combining and correlating information from all sub-detectors. The CMS detector design is well-suited to the use of PF reconstruction: a highly-segmented tracker, fine-grained electromagnetic calorimeter, hermetic hadron calorimeter, strong magnetic field and excellent muon spectrometer provide full, high quality coverage~\cite{sirunyan2017pflowcms}. The ATLAS experiment implements PF algorithms offline and partly online. The ATLAS detector gains less from the approach due to its excellent energy resolution in the calorimeters but for low-momentum resolution and pile-up rejection the improvement is substantial~\cite{ATLASTriggerRun3,ATLASJetPFlow}.

\subsection{ML jet and $b-$jet identification algorithms in ATLAS} \label{sec:Algorithms_bTag}

More recently, machine-learning-based algorithms have emerged as options for fast reconstruction. In particular, Graph Neural Networks (GNNs), a class of machine learning models designed to operate on graph-structured data, have seen increasing adoption across the LHC experiments. For example, GNNs are used in the ATLAS Run~3 $b$-jet trigger, where tracks of common vertices are grouped and predictions around jet origin are made~\cite{ATLASTriggerRun3}. 

GNNs propagate information through the nodes and edges of a graph, capturing complex relationships and dependencies within the data. GNNs can learn to perform various tasks, such as node classification, link prediction, and graph classification, for example the cluster of hits and reconstuction of showers in the LHCb calorimeter systems~\cite{canudas2022graph}.

\subsection{TPC cluster finding in ALICE}

Online TPC cluster and track finding algorithms are amongst the most computationally intensive processes of the ALICE TPC, as described in Ref.~\cite{alice-rta-trigger}.
Both algorithms can be designed to be highly-parallisable and hence portable to heterogeneous architectures.
For example, the TPC cluster finding algorithm which consists of three stages: signal extraction/calibration, identification/center of gravity calculation of neighbouring signals, and merging of neighbouring clusters between TPC readout rows.
Since each stage consists of small operations with low memory requirements, this algorithm is ideally suited to run on FPGAs~\cite{alice:fpga}.

\section{Alignment and calibration of detectors}

Accurate alignment and calibration of detectors are crucial to the quality of event reconstruction, ensuring faithful decoding of raw detector information to reconstructible objects. This is particularly important for offline-quality reconstruction, where the final reconstructed objects are later used in physics analysis. Alignment and calibration are typically implemented together %(where alignment often refers to tracking subdetectors and calibration to non-tracking subdetectors, e.g., calorimeters),
performing a minimisation of parameters describing the physical offsets experienced in a given subdetector, e.g., residuals between the expected position of a track and the measured position. Alignment and calibration techniques are also applied online when partial reconstructions and trigger decisions require more accurate knowledge of an object. The real-time approaches of LHCb and ALICE in Run~3 typify this and are thus the focus of this chapter. 
At ATLAS and CMS, the more complex detector nature and the need for the full dataset to obtain sufficiently large calibration samples currently hinders a similar full alignment and calibration strategy in real time.

% LHCb
%\subsection{Real-time detector alignment in LHCb}
To enable the use of real-time alignment and calibration of the LHCb detector the trigger system was designed with a large timing budget between HLT stages, provided by a disk buffer of 30 PB. The online LHCb calibration ensures offline-quality reconstruction, enabling trigger selection with high signal efficiencies. This calibration is typically separated into the alignment of the VELO, UT/SciFi, muon stations and RICH mirrors, and the calibration of the ECAL and RICH. A dedicated data sample\footnote{In addition to triggers which select events for offline analysis at LHCb, dedicated triggers which select events required for alignment and calibration purposes are employed. Events selected by the latter are saved into a dedicated stream described in Section~\ref{sec:data}.} is collected, from which the alignment framework calculates updated alignment constants at regular intervals, i.e., per fill or run, as soon as sufficient data has been collected for precise calculation of the constants.

The alignment constants for the tracking detectors are determined by minimising track reconstruction parameters with respect to the degrees of freedom of each alignable detector element, i.e., translations and rotations in each spatial dimension~\cite{LHCb:wouter-kf}. The UT and SciFi are also aligned using reconstructed tracks traversing the entire tracking system to achieve high track-momentum resolution~\cite{Reiss:2846414}. The ECAL is calibrated by analysing a specific mass distribution in each ECAL cell, providing a mass shift from the known position which can be applied as an adjustment of the photomultiplier tubes after each fill. The RICH is aligned and calibrated on a per-run basis, comparing the distribution of hits and Cherenkov angles to anticipated distributions to generate alignment and calibration constants~\cite{LHCb:RICH_AlignCalib}.
%the $\pi^0\rightarrow \gamma\gamma$ 

% ALICE
%\subsection{Real-time TPC alignment in ALICE}
At ALICE, as a consequence of upgrades for continuous readout in Run~3, the TPC requires real-time calibration to correct for space-charge density distortions which are particularly problematic in high pile-up scenarios. Fluctuations in space-charge distortions vary on a scale of \SI{10}{\milli\second} and need to be corrected for. Average space-charge distortions are corrected with the aid of relevant subdetectors; however, the calibration of average space-charge distortions requires $\mathcal{O}\left(\mathrm{minutes}\right)$ of data collection to form a reliable correction map. Convolutional neural networks (CNNs) are employed to predict the fast-varying fluctuations in space-charge distortion, obtaining a reliable calibration at the fluctuation timescale. The CNN models run on GPUs directly in the HLT farm, resulting in a significant speedup of the TPC calibration.

\section{Processing and storage of data}
\label{sec:data}

Reconstructed detector events are typically complex data objects described by many parameters. 
% The size of events presents a storage challenge as it can become infeasible to write all raw/reconstructed data to storage at the output rate of the high-level trigger. 
The size of full events presents a storage challenge and the HLT output rate is the maximum limit at which raw and reconstructed data can be written to disk.
To overcome this challenge, two key strategies are employed to reduce the size of the reconstructed objects. Firstly, the amount of information can be reduced, by storing only the minimal set of observables necessary to describe the process of interest. Secondly, the object information itself can be compressed, maintaining most or all of the original information while significantly reducing the size of the data being written.

% Streaming
%\subsubsection{Data scouting at ATLAS (TLA) and CMS}
An example of the first approach is the CMS data scouting stream (akin to trigger level analysis (TLA) in ATLAS~\cite{ATLAS:TLA} and Turbo stream in LHCb~\cite{Aaij:2019uij}), wherein events are stored at higher rates with smaller event content - storing simply the online objects, bypassing offline reconstruction entirely~\cite{ardino202340cms,tomei2020cms,badaro202040cms}. 
In addition, a parking stream or delayed stream saves event data without running offline reconstruction algorithms, with the intent of processing said events during shutdown periods when computing resources are less constrained. 
CMS monitoring and calibration streams are also able to make use of their small event size to store more events, with typical calibration stream event sizes in 2022 and early 2023 data-taking of $\SI{13}{\kilo\byte}$, in comparison to an event with full detector readout at around \SI{1}{\mega\byte}~\cite{CMS:run3-detector}. However, the additional information of the latter event model allows a wide range of track reconstruction algorithms to be applied for the identification of $b$-jets, lepton isolation and the mitigation of pile-up vertices~\cite{tosi2016trackingcms}. 

\begin{figure}[!ht]
    \centering
    \includegraphics[width=0.55\linewidth]{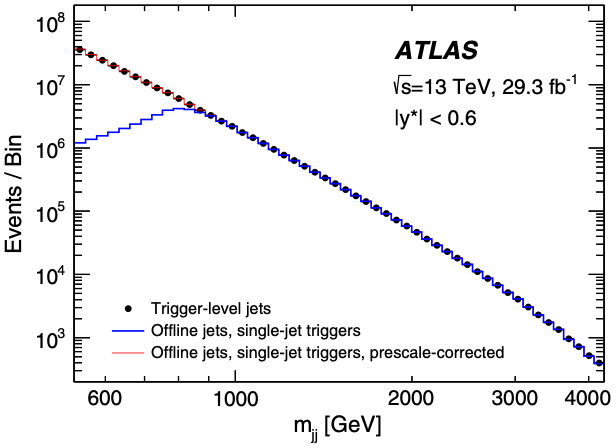}
    \caption{Comparison of offline and trigger-level jet objects in ATLAS dijet spectra from Ref.~\cite{ATLAS:2018qto}. Trigger-level jets allow recording low $m_{jj}$ events at the full rate by greatly reducing the throughput.}
    \label{fig:trigger-level-efficiencies}
\end{figure}

%\subsubsection{The LHCb Turbo model}
In the LHCb Turbo model, only the objects relevant to an HLT2 trigger decision are persisted and saved to a dedicated stream~\cite{Aaij:2016rxn}.
The reduced event size thus allows to record event at a higher rate. Many trigger-level objects may be required which are beyond the standard objects of a given decay, for example for flavour tagging. 
In such cases, selective persistency is applied, whereby an event is saved in the Turbo model with specified additional information, e.g., other tracks arising from a primary vertex, objects contained in a cone around the candidate, etc., A calibration stream (TurCaL) is a special use case of the selective persistence model, wherein only events dedicated to detector alignment and calibration are persisted~\cite{Aaij:2019uij}.

A similar approach is taken in ATLAS with the Partial Event Building (PEB) stream of events containing only partial detector information~\cite{ATLASTriggerRun3}. These reduced-size events enable dedicated higher-rate triggers to be run in the PEB stream, such as calibration and jet triggers.
\section{Real-time analysis in physics results}
\label{sec:RTA_physics}
%Caterina having a first go, but students will edit further during circulation period

In this section, we briefly outline and give references for some of the physics analyses that have been performed by ATLAS, CMS and LHCb using real-time analysis (RTA) in Run~1 and Run~2. The success of such analyses has further motivated the development and adoption of RTA techniques in Run~3 trigger systems. %, as summarised in Table~\ref{trigger-table}.

The data scouting stream in CMS has been in place since the LHC Run~1, and its first use in searches for dijet resonances is described in Ref.~\cite{CMS:dijet-7tev-resonances} and a review of all CMS scouting searches can be found in Ref.~\cite{CMS:2024zhe}. The RTA approach allows the reach of this search to be extended to resonances with masses as low as \SI{500}{\giga\electronvolt} improving on the standard data-taking analysis that would only be sensitive to resonances with masses of 1 TeV and above\footnote{Other techniques also exist beyond RTA to reach lower resonance masses, e.g., by triggering on initial state radiation or considering boosted jets that contain the resonance products.}. The same search has also been performed with the 8 and 13 TeV centre-of-mass LHC datasets in Run~1 and Run~2, and described in Refs.~\cite{CMS:2016ltu, CMS:2016gsl}, extending the reach to 500 GeV resonances. ATLAS also searched for dijet resonances using the TLA technique, with sensitivity for resonances with masses as low as 450 GeV, as described in Ref.~\cite{ATLAS:2018qto}.

%CD: missing dijet+ISR? 

Searches for multijet resonances (new particles decaying in paired dijet and three jets each) also employ, for example in Ref.~\cite{CMS:scouting-search-multijet}. RTA extends the reach of previous searches to the low mass region from 200 GeV down to 70 GeV for RPV squarks and gluinos. 

Long-lived particles (LLPs) predicted by beyond-SM models decay far from the point of collision, a signature distinct from the promptly decaying particles of the majority of LHC searches. LLPs are often rejected by standard reconstruction algorithms due to their unusual characteristics (e.g., display decay vertices)~\cite{llps}. As such, LLPs require dedicated selection techniques and searches, for example, in CMS real-time searches for dark photons and LLPs decaying into muons using Run~2 data.  Ref.~\cite{CMS:2019buh} describes a search for dark photons decaying into dimuons that uses RTA in the 11.5-45 GeV mass range. In Ref.~\cite{CMS:2021sch}, RTA enables access to the new phase space of low dimuon masses and non-zero displacement that would otherwise not be covered by standard searches.

%\subsection{ATLAS}

%\subsection{LHCb}

In LHCb, real-time analysis was motivated~\cite{Gligorov:2018fuk} by the size of the LHC production cross-section of hadrons containing a charm 
quark. Because this cross-section is so large, it is not possible to record all signal decays of charmed hadrons to permanent storage while keeping the full detector information for each event. Therefore, these decays must be fully reconstructed and selected in real-time, requiring accurate and up-to-date detector alignment and calibration in the real-time processing to keep systematic uncertainties under control. This is particularly crucial since LHCb has a unique ability to probe CP violation in charm hadrons to the $10^{-5}$ level or better, requiring a corresponding control of systematics. For this reason, LHCb implemented the full offline-quality reconstruction, alignment, and calibration~\cite{Dujany:2015lxd, Aaij:2016rxn, Borghi:2017hfp, LHCb:2018zdd, Aaij:2019uij} of the detector within its real-time processing (specifically the software trigger) during Run~2, and adopted real-time analysis as the baseline model~\cite{LHCbCollaboration:2319756} for the majority of the collaboration's physics programme from Run~3 onwards. In Run~2, almost all analyses of charm hadrons, as well as certain searches for BSM states (most notably dark photons), were carried out using real-time analysis.
\section{Outlook on HL-LHC Upgrade}

% ATLAS+CMS to 1 level
% LHCb timing info?
% ALICE ???
% Possible approaches
% TDRs?

Following the conclusion of Run~3 (expected 2025), the LHC will enter the HL-LHC operational phase (expected 2029), with the installation of many detector upgrades in the interval. Accelerator upgrades will increase the LHC instantaneous luminosity up to \SI{7.5e34}{\per\square\cm\per\second}, a factor ten increase on the design luminosity. Access to this increased luminosity will allow the experiments to collect datasets an order of magnitude larger. This enormous increase in sample size enables the experiments to push both the precision and intensity frontiers of research.

The HL-LHC operation poses a series of challenges to the continued efficient operation of the trigger systems of the LHC experiments. For example, in ATLAS and CMS, pile-up is likely to increase to between 140 and 200 proton-proton interactions~\cite{ATLAS:pileup} from an average value of 55 in Run~3. As the occupancy and complexity of each event increases, so will the processing time required to make effective trigger decisions. %Increasing L1 and HLT trigger rates? here
The increase in event complexity brings an associated expansion in the memory footprint/bandwidth requirements. 
Furthermore, since QCD cross-sections scale linearly with pile-up, bandwidth limitations will result in significantly reduced fractions of low-energy jet selection, in turn hindering the sensitivity of low-energy jet searches~\cite{albrecht2018hep}.

In ATLAS and CMS, several improvements and innovations in both detector/hardware and software are under development targeting HL-LHC~\cite{hl-lhc}.

The upgrades to the sub-detectors will improve trigger performance (e.g., MIP/HGCAL in CMS~\cite{cms2019mip, cms2017phase-hgcal}, Calo~\cite{ATLAS:ECAL, ATLAS:HCAL}/Muon~\cite{ATLAS:Muon} in ATLAS L1, HGTD~\cite{ATLAS:HGTD}/ITK~\cite{ATLAS:ITKPixel, ATLAS:ITKStrip} in ATLAS HLT), or permit previously inaccessible combinations of sub-detectors (e.g., New Tracker in CMS~\cite{collaboration2017phasecms}). Additionally, upgrades to onboard electronics (e.g., FELIX in ATLAS~\cite{ATLAS:FELIX, ATLAS:TDAQ}) will handle the increased readout requirements as well as the increased complexity of events by allowing an increased L1 latency. Such increases in readout capacity will also be of vital importance as new/upgraded subdetectors will further increase readout requirements (e.g., HGCAL in CMS~\cite{cms2017phase-hgcal}).
In both experiments, there is an intense focus on heterogeneous computing architectures and their application in the real-time processing of data~\cite{ATLAS:c-and-s-roadmap,bocci2020heterogeneouscms}. ASICs and FPGAs, traditionally used in the initial stages of the data pipeline, are being deployed more widely and in diverse use cases. Migration of parallelisable tasks to GPUs is projected to reduce the power consumption, computational cost and wall time of the present trigger menus~\cite{cms-GPU-clustering}. 

Current software trigger algorithms have also been subject to scrutiny and the developments here are in many cases quicker to iterate than hardware upgrades. All of the algorithmic improvements would be too numerous to list here. We include just the following examples for brevity.
The ATLAS experiment intends to use novel GNN models to reduce resource consumption in track reconstruction~\cite{Caillou:2815578}, to be deployed on dedicated GPU cores in the new HLT.
In the CMS experiment the PF algorithm, described in Section~\ref{sec:Algorithms_PFlow}, will be launched inside the new L1 trigger in Run~4, and a correlator trigger will make use of tracks that are read out at \SI{40}{MHz} for the first time. A pile-up mitigation algorithm (PUPPI) will be used in L1 - requiring primary vertex identification.

In the HL-LHC paradigm, the use of specialised data streams based on reduced event content—Data Scouting in CMS~\cite{ardino202340cms,tomei2020cms,badaro202040cms}, TLA in ATLAS~\cite{ATLAS:TLA}, Turbo in LHCb~\cite{Aaij:2019uij}—will be expanded.
% NOT HAPPY WITH PREV PARA

%CMS
%https://cds.cern.ch/record/2714892
%https://arxiv.org/pdf/2010.13557.pdf %CMS track in trigger
%atlas
%https://cds.cern.ch/record/2285584

The LHCb and ALICE experiments will not make fundamental changes to their current running strategy for Run~4. Rather, major upgrades are scheduled for LS4 in preparation for Run~5 (to commence in 2035)~\cite{CERN-LHCC-2021-012, alice_loi_hl_lhc}.

\section{Conclusion}

The trigger systems at the major LHC experiments have undergone a continuous evolution throughout their operation in Run~1 and Run~2. Significant upgrades have been implemented and proposed to address the challenges posed by the current Run~3 and future high luminosity conditions. The triggers now take advantage of modern hardware developments in the sub-detector readouts and novel data processing software including contemporary machine learning developments. The result today is a more robust, flexible and efficient trigger system, better able to accommodate future requirements.

In Run~3, the major LHC experiments leverage their experience from Run~1 and Run~2 to tackle increased data challenges. In particular, upgrades to these experiments and their respective trigger systems take advantage of recent advances in hardware technologies, with software frameworks redesigned to better capitalise on such upgrades. New and optimised software frameworks also provide implementations for faster and more capable algorithms for reconstruction, selection and data manipulation. Overall, this experience will be integrated into the plans for Run-4, also taking advantage of new detectors and new trigger hardware capabilities.

\clearpage
\section*{Acknowledgements}
%This is important, as it acknowledges our funding. 
This work is part of the SMARTHEP network and it is funded by the European Union’s Horizon 2020 research and innovation programme, call H2020-MSCA-ITN-2020, under Grant Agreement n. 956086. 
We thank Claire Antel, Gerhard Raven and Michael David Sokoloff for providing useful comments to this whitepaper.

\bibliography{references, references_intro, references_LHCb, references_ALICE, reference_ATLAS, references_CMS, references_RTA, references_HLLHC}
\bibliographystyle{JHEP}

\end{document}